\PassOptionsToPackage{numbers,sort&compress}{natbib}

\documentclass[twocolumn,pdflatex,sn-nature]{sn-jnl}

\usepackage{amsmath, amssymb, bm}
\usepackage{graphicx}
\usepackage{booktabs}
\usepackage[font=footnotesize,labelfont=bf]{caption}
\usepackage{xcolor}
\definecolor{assistgray}{gray}{0.6}
\usepackage[labelfont=bf,labelsep=period]{caption} 
\renewcommand{\figurename}{Figure}
\renewcommand{\tablename}{Table}
\renewcommand{\abstractname}{Abstract}

\newcommand{\longcomment}[1]{} 

\usepackage{setspace}

\usepackage[utf8]{inputenc}

\usepackage{multirow} 
\usepackage{makecell}

\usepackage{microtype}  

\usepackage{threeparttable}   
\usepackage{url}              

\usepackage{placeins} 

\usepackage{float} 

\usepackage{afterpage}

\usepackage{pdflscape}  

\usepackage{booktabs}

\graphicspath{{Fig/}}

\theoremstyle{thmstyleone}%
\theoremstyle{thmstyletwo}%
\theoremstyle{thmstylethree}%

\begin{document}

\sloppy   

\title[Short Article Title]{zERExtractor:An Automated Platform for
Enzyme-Catalyzed Reaction Data Extraction from
Scientific Literature}

\author[1,2]{\fnm{Rui} \sur{Zhou}}
\equalcont{These authors contributed equally to this work.}
\author[3]{\fnm{Haohui} \sur{Ma}}
\equalcont{These authors contributed equally to this work.}

\author[4]{\fnm{Tianle} \sur{Xin}}
\author[5]{\fnm{Lixin} \sur{Zou}}
\author[1,2]{\fnm{Qiuyue} \sur{Hu}}
\author[5]{\fnm{Hongxi} \sur{Cheng}}
\author[5,3]{\fnm{Mingzhi} \sur{Lin}}
\author[6]{Jingjing Guo}
\author[3]{\fnm{Sheng} \sur{Wang}}
\author*[4,7]{\fnm{Guoqing} \sur{Zhang}} \email{gqzhang@sinh.ac.cn}
\author*[1]{\fnm{Yanjie} \sur{Wei}} \email{yj.wei@siat.ac.cn}
\author*[5,3]{\fnm{Liangzhen} \sur{Zheng}} \email{zhenglz@zelixir.com}



\affil[1]{\orgdiv{Shenzhen Institutes of Advanced Technology}, \orgname{Chinese Academy of Sciences}, \city{Shenzhen}, \postcode{518000}, \country{China}}

\affil[2]{\orgdiv{University of Chinese Academy of Sciences}, \city{Beijing}, \postcode{100049}, \country{China}}

\affil[3]{\orgdiv{Shanghai Zelixir Biotech Co.,Ltd.}, \city{Shanghai}, \postcode{201203}, \country{China}}

\affil[4]{\orgdiv{Bio-Med Big Data Center, Shanghai Institute of Nutrition and Health, University of Chinese Academy of Sciences, Chinese Academy of Science}, \city{Shanghai}, \postcode{200031}, \country{China}}

\affil[5]{\orgdiv{Shenzhen Zelixir Biotech Co., Ltd.}, \city{Shenzhen}, \postcode{518107}, \country{China}}

\affil[6]{\orgdiv{Centre in Artificial Intelligence Driven Drug Discovery, Faculty of Applied Sciences, Macao Polytechnic University}, \orgaddress{\city{Macao SAR}, \country{China}}}

\affil[7]{\orgdiv{National Genomics Data Center, Shanghai Institute of Nutrition and Health, University of Chinese Academy of Sciences, Chinese Academy of Science}, \city{Shanghai}, \postcode{200031}, \country{China}}



\abstract{
The rapid expansion of enzyme kinetics literature has outpaced the curation capabilities of major biochemical databases, creating a substantial barrier to AI-driven modeling and knowledge discovery. We present zERExtractor, an automated and extensible platform for comprehensive extraction of enzyme-catalyzed reaction and activity data from scientific literature. zERExtractor features a unified, modular architecture that supports plug-and-play integration of state-of-the-art models—including large language models (LLMs)—as interchangeable components, enabling continuous system evolution alongside advances in AI. Our pipeline synergistically combines domain-adapted deep learning, advanced OCR, semantic entity recognition, and prompt-driven LLM modules, together with human expert corrections, to extract kinetic parameters (e.g., $\text{k}_\text{cat}$, $\text{K}_\text{m}$), enzyme sequences, substrate SMILES, experimental conditions, and molecular diagrams from heterogeneous document formats.
In this framework, active learning that tightly integrates AI-assisted annotation, expert validation, and iterative model refinement—allowing the system to adapt rapidly to new data sources. We also curate and release a large, expert-annotated benchmark dataset comprising over 1,000 fully annotated tables and 5,000 biological fields from 270 peer-reviewed P450-related enzymology publications, establishing a valuable and extensible resource for the enzyme informatics community. 
Benchmarking demonstrates that zERExtractor consistently outperforms existing baselines in table recognition (Acc 89.9\%), molecular image interpretation (up to 99.1\%), and relation extraction (role assignment accuracy 94.2\%). By bridging the longstanding data gap in enzyme kinetics with a flexible, plugin-ready framework and high-fidelity extraction, zERExtractor lays the groundwork for future AI-powered enzyme modeling and biochemical knowledge discovery.
}
\maketitle

\section{Introduction}

Experimentally measured enzyme kinetic data are essential for uncovering the catalytic origin of enzymatic function and evolution, while also serving as the foundation for training predictive AI models of enzymatic rate constants and specificity constants—a field that has witnessed remarkable advances in recent years\cite{li2022deep_dlkcat,wang2025robust_catapro,yu2023unikp}.
The quality and quantity of available enzyme kinetic data critically influence the progress of research on predictive modeling of enzyme kinetic parameters\cite{kroll2023turnover}.

\begin{figure}[htbp]
  \centering
  \includegraphics[width=1.0\linewidth]{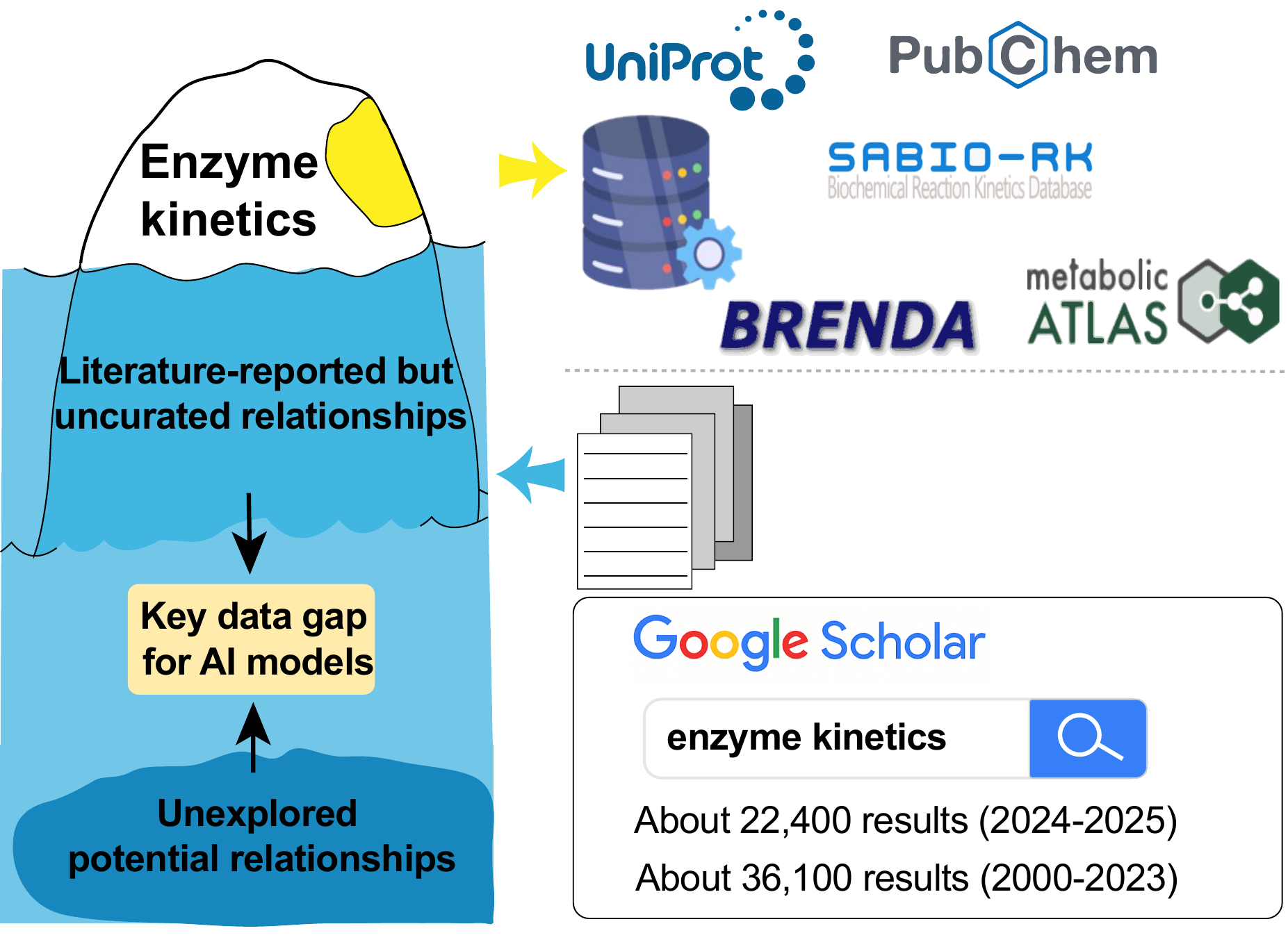}
  \caption{{Multilayered Data Landscape and Curation Bottleneck in Enzyme Kinetics Research.}A schematic view of the enzyme kinetics data landscape highlights the ``iceberg effect'': only the curated tip is included in public databases (e.g., BRENDA, SABIO-RK, UniProt, PubChem, Metabolic Atlas\cite{pornputtapong2015human}), while most literature-reported data remain uncurated and inaccessible to AI models. The yellow region at the iceberg tip represents data that meet the curation standards required by public biochemical databases.At the base lies a vast unexplored space of enzyme–substrate–condition combinations. The right side shows the gap between reported studies and database coverage.}
  \label{fig_introData}
\end{figure}

In addition to well-curated databases such as BRENDA\cite{chang2021brenda,schomburg2017brenda} and SABIO-RK~\cite{wittig2018sabio}, large-scale biological repositories including deep mutational scanning (DMS) datasets~\cite{fowler2014deep}, UniProt~\cite{uniprot2023uniprot}, and PubChem\cite{kim2021pubchem} offer extensive information on protein sequences, functional variants, and small molecule substrates. These resources collectively provide rich contextual data for modeling enzyme–substrate interactions and kinetic behaviors. Nevertheless, they encompass only a limited portion of the kinetic parameters described across the scientific literature, resulting in a substantial volume of sequence–function–condition relationships remaining uncurated and underexplored\cite{pleiss2024modeling}.
Meanwhile, the scientific output in enzyme kinetics as well as enzymatic reactions has been growing at an unprecedented pace, with thousands of new publications appearing each year and the cumulative body of literature now numbering in the hundreds of thousands.
As illustrated in  Figure~\ref{fig_introData},only a small fraction of enzyme kinetics data reported in the literature is curated in major biochemical databases, leaving the majority uncurated or unexplored.
This rapid expansion far exceeds the capacity of existing databases to systematically curate kinetic data, highlighting a critical gap between the wealth of information embedded in publications and the accessible, structured data available for computational modeling and machine learning applications.

Recent advancements in Artificial General Intelligence (AGI), particularly in LLMs, have shown remarkable capabilities in mining scientific literature, including named entity recognition, relation extraction, summarization, information integration, and even scientific planning~\cite{hu2025evaluation,jin2024pubmed,dagdelen2024structured,polak2024extracting,boiko2023autonomous}.
In the context of enzyme engineering, recent efforts have emerged that leverage AI-assisted approaches to extract enzymology-related data—such as kinetic parameters, substrates, and experimental conditions—from the vast body of biochemical literature, aiming to accelerate data acquisition for downstream modeling and design\cite{smith2025funcfetch,lai2024enzchemred,jiang2025enzyme}.

In contrast, our work introduces zERExtractor, a unified and scalable framework designed to overcome these bottlenecks. By integrating domain-adapted deep learning with LLMs in a human-in-the-loop pipeline, zERExtractor enables accurate extraction and structuring of enzyme-catalyzed reaction data across diverse document formats. Unlike previous systems, it captures not only structured kinetic parameters (e.g., $k_\text{cat}$, $K_\text{m}$), but also enzyme sequences, relative activities, yields, substrate/product SMILES, experimental metadata, and molecular structure diagrams. We further construct and release a large-scale, expert-annotated benchmark dataset covering 270 P450-related open-accessed publications, supporting both training and evaluation. Quantitative evaluations show that zERExtractor consistently outperforms existing baselines in table recognition, molecular image interpretation, and reaction extraction. Together, these advances bridge a longstanding gap between literature and structured biochemical knowledge, laying a robust foundation for AI-driven enzyme function modeling and design.


In summary, our main contributions are as follows:\\
\textbf{A unified multimodal extraction system:} We propose zERExtractor, an automated and scalable platform that integrates domain-adapted deep learning and LLMs to extract enzyme-catalyzed reaction data—including kinetic parameters, enzyme sequences, substrate SMILES, experimental conditions, and molecular diagrams—from unstructured scientific literature.

\textbf{Adaptive model improvement through expert-guided active learning:}We implement a feedback-driven pipeline that combines AI-assisted annotation with manual validation and active sample selection, enabling continuous model refinement, improved robustness, and adaptive performance across new data sources.

\textbf{A high-quality expert-annotated benchmark dataset:}We curate and release a large-scale dataset of over 1,000 fully annotated tables and approximately 5,000 biological fields from 270 peer-reviewed enzymology publications, providing a valuable resource for model training, evaluation, and benchmarking in the enzyme informatics community.

\section{Method}

\begin{figure*}[htbp]
    \centering
    \includegraphics[width=\linewidth]{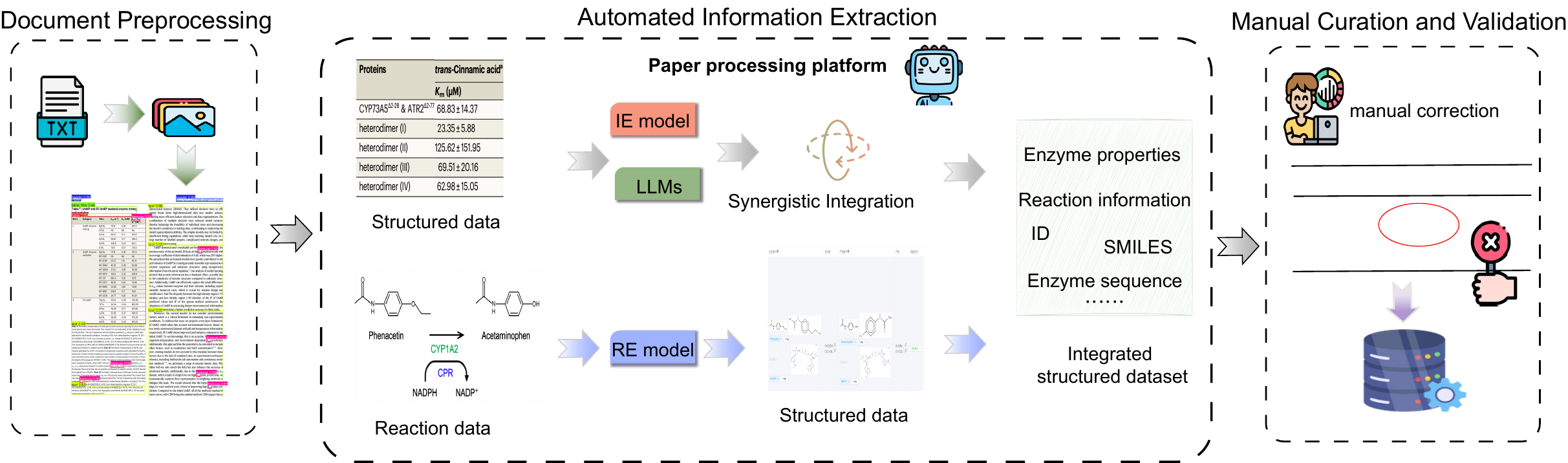}
    \caption{{ A pipeline specialized in enzyme-catalyzed reaction information extraction.}}
    \label{fig_overview}
\end{figure*}

\subsection{A pipeline specialized in enzyme-catalyzed reaction information extraction}
We developed a comprehensive and modular pipeline for the extraction of enzyme-catalyzed reaction information, which seamlessly integrates both advanced deep learning techniques and LLMs. As illustrated in Figure~\ref{fig_overview}, the workflow is designed to address the unique challenges posed by biochemical literature, where critical data are dispersed across tables, molecular diagrams, and unstructured text.

The pipeline begins with the ingestion and parsing of raw PDF documents, which are segmented into figures, tables, and textual content. For the table recognition module, we employ a hybrid approach that combines state-of-the-art object detection algorithms (such as PP-YOLOv2 and SLANet) 
~\cite{redmon2017yolo9000,li2022pp}
with prompt-driven LLMs. The object detection models are responsible for accurately locating table regions in complex document layouts, while the LLMs enable robust schema alignment and semantic interpretation of extracted table content, even in the presence of multi-line headers, merged cells, and domain-specific terminology.
Similar to prior formulations that leverage optimization-based contact prediction in structured biological contexts~\cite{fan2024openchemie}, our pipeline combines learning and constraint-based alignment to improve structural fidelity.

For molecular structure recognition, we introduce UniMolRec, a unified recognition framework that we developed specifically for this study. UniMolRec integrates multiple deep learning models in an ensemble fashion to robustly convert molecular structure images into canonical SMILES representations, ensuring high performance and adaptability across diverse image styles and noise conditions.

A key feature of our system is the incorporation of an interactive annotation and validation module, which enables human experts to review and refine automatically extracted data\cite{monarch2021human}. The validated and corrected annotations are systematically reintegrated into the model training loop, thereby allowing the underlying algorithms to continuously improve in both accuracy and robustness. This human-in-the-loop framework ensures that the system can adapt to new document formats and data patterns, and that high-quality, semantically accurate structured data are maintained over time.

The pipeline further incorporates LLM-based modules for entity and relationship extraction, which accurately assign roles such as substrate, product, enzyme, and cofactor, and facilitate cross-modal semantic integration. A human-in-the-loop annotation workflow, augmented by an active learning module, allows for continuous refinement of extraction models and ensures high data reliability\cite{monarch2021human}.

By synergistically combining deep learning and LLM capabilities, this hybrid pipeline significantly improves the efficiency, accuracy, and generalizability of enzyme-catalyzed reaction information extraction. The resulting structured datasets serve as a robust resource for downstream applications in computational enzymology and AI-driven biochemical analysis\cite{li2022deep_dlkcat,wang2025robust_catapro,kroll2023turnover,luo2024benchmarking,yu2023unikp}.

\subsubsection{Automated Extraction of Reaction Schematics}

Molecular structure recognition involves translating single-molecule images into their corresponding chemical structures, serving as a crucial step in digitizing chemical information from scientific literature\cite{clevert2021img2mol,khokhlov2022image2smiles,hirohara2018convolutional}. Recent approaches have leveraged convolutional neural networks and transformer-based architectures to directly map molecular depictions to SMILES or other chemical representations, enabling automated and scalable interpretation of structural diagrams across diverse document sources\cite{hirohara2018convolutional,khokhlov2022image2smiles}.

We developed UniMolRec, a unified recognition framework designed for molecular structure interpretation. By integrating multiple specialized models through an ensemble architecture, UniMolRec achieves robust and accurate translation of molecular images into canonical chemical representations.

While single-model approaches often suffer from limited generalizability across diverse molecular image styles and noise conditions, we hypothesize that an ensemble strategy allows us to combine complementary strengths of multiple models, leading to more robust and accurate recognition\cite{ju2018relative}.

\begin{figure*}[htbp]
    \centering
    \includegraphics[width=\linewidth]{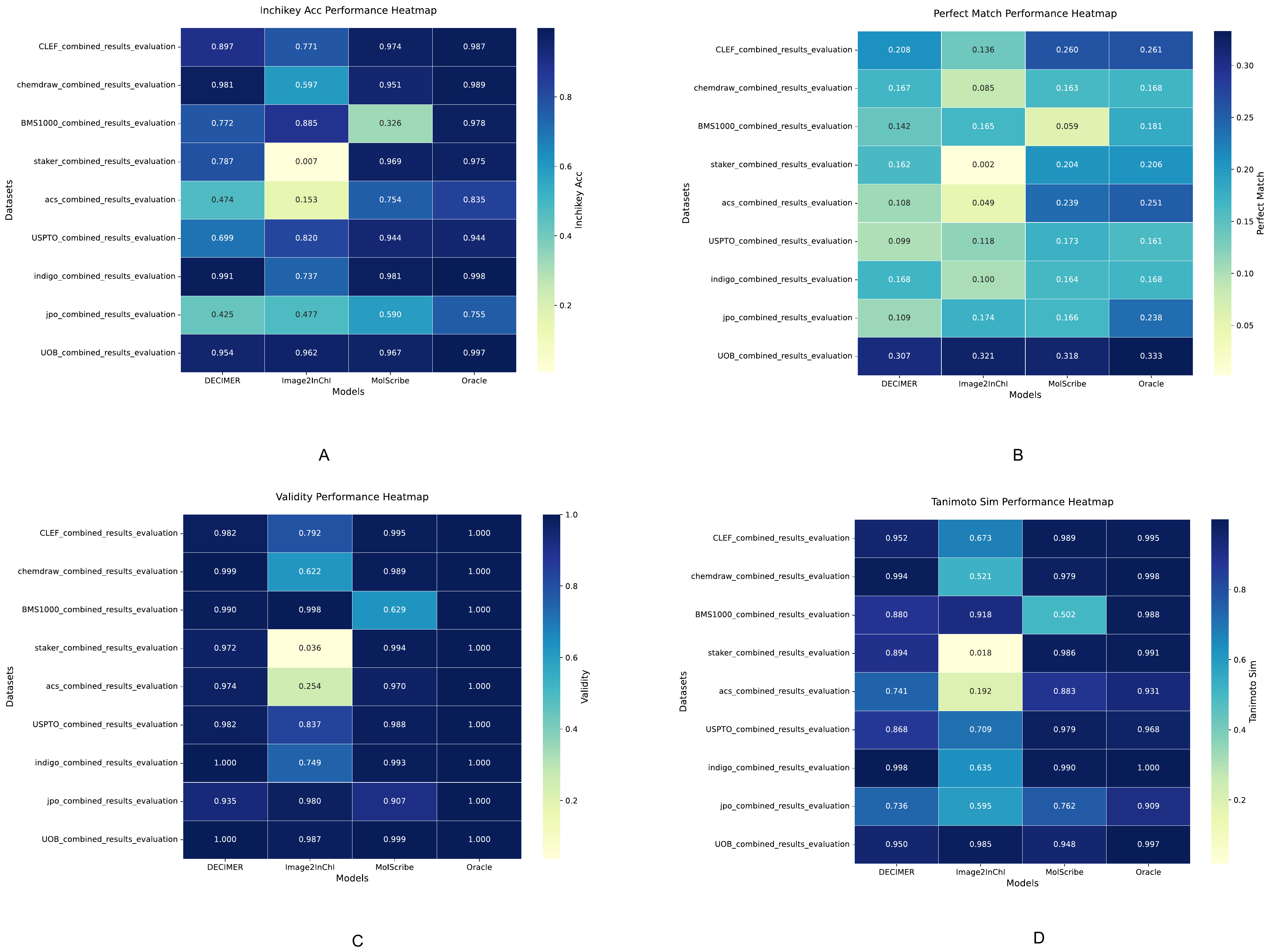}
    \caption{{Multi-metric performance heatmaps comparing different molecular recognition models across multiple benchmark datasets.}
    \textbf{Panel A} shows the InChIKey Accuracy, indicating the correctness of the predicted molecule’s standardized chemical identifier.
    \textbf{Panel B} shows the Perfect Match ratio, reflecting the percentage of predictions that exactly match the ground truth structure.
    \textbf{Panel C} reflects the proportion of chemically valid molecules generated by each model.
    \textbf{Panel D} displays the Tanimoto Similarity, measuring structural similarity based on molecular fingerprints.
    }
    \label{fig_headmap}
\end{figure*}

\begin{figure*}[htbp]
    \centering
    \includegraphics[width=\linewidth]{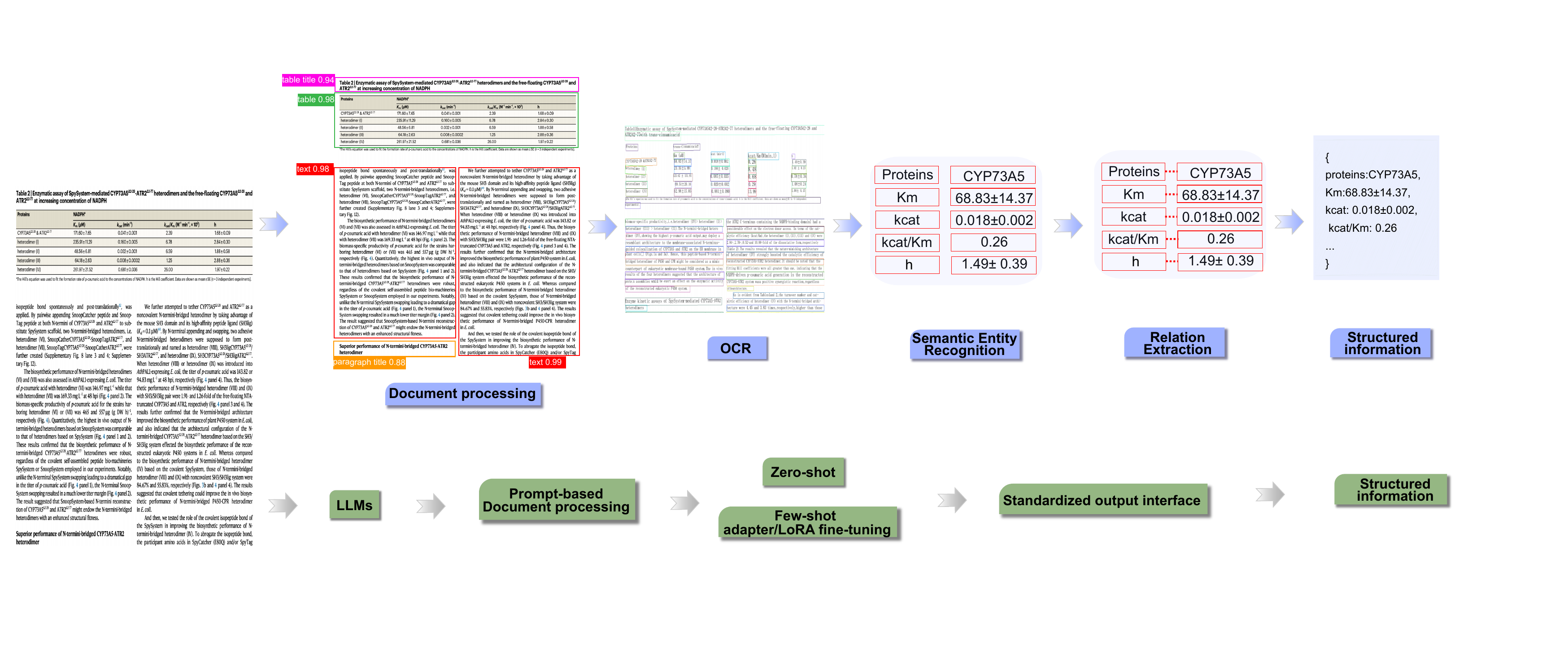}
    \caption{{ Multi-Module Workflow for Structured Extraction of Enzyme Kinetic Parameters.}(Top) Traditional pipeline based on deep learning and OCR. Raw PDFs are first processed to identify table regions, followed by OCR to extract layout-aware cell content. Semantic entity recognition and relation extraction modules are applied to identify key biological fields such as enzyme name, $K_m$, $k_\text{cat}$, and pH, and align them into a structured schema. The final output is converted into JSON format with value-unit pairs.
    (Bottom) LLM-based document understanding pipeline. Prompt-based document parsing is combined with zero-shot or few-shot tuning (e.g., via LoRA) to directly extract structured biochemical entries from unstructured text or semi-structured tables. This pathway simplifies pipeline complexity while enhancing generalization across diverse document formats.}
    \label{fig_tableGet}
\end{figure*}

As illustrated in Figure~\ref{fig_headmap}, although many state-of-the-art models\cite{qian2023molscribe,rajan2023decimer} achieve high performance on specific datasets, their generalization capability across diverse molecular image distributions remains limited.

Effective extraction of small molecules from PDFs requires careful consideration of model generalizability across varied document layouts and image qualities. Our proposed framework, UniMolRec, demonstrates strong performance and robust adaptability under these challenging conditions.

\subsubsection{Structured Extraction of Tabular Data}

We collected 270 open-source scientific articles related to cytochrome P450 enzymes\cite{bernhardt2006cytochromes_p450,gillam2008engineering_p450,shaik2010p450_p450}, with a focus on extracting enzymatic reaction data. These articles, spanning from 2000 to 2024, primarily belong to the fields of biochemistry, structural biology, and synthetic biology. To ensure relevance to our study, we filtered the documents based on keyword matching, retaining only those containing at least one experimental table. Particular attention was given to tables reporting enzyme kinetic parameters such as $k_\text{cat}$ and $K_\text{m}$, as well as reaction-related information including substrate SMILES and enzyme sequences. These curated documents served as the raw material for the subsequent tabular data extraction pipeline.

To extract tabular data from the curated PDFs, we employed PP-Structure, a structure-aware OCR system developed within the PaddleOCR framework\cite{li2022pp}. This model was fine-tuned on domain-specific samples to better accommodate the characteristics of enzyme kinetics tables, including multi-line headers, merged cells, and nested parameter groupings. The extraction process leveraged both textual and visual layout cues—such as cell bounding boxes and spatial relationships—to reconstruct table structures accurately.
The extracted tables were then converted into structured JSON files, where each row corresponds to a biological reaction entry. The schema captures domain-specific fields such as Enzyme name, Substrate, Enzyme ID (e.g., UniProt, GenBank, or PDB), Product, pH, $K_m$, $k_\text{cat}$, Mutations, Source Organism, and Relative Activity. Each quantitative field was extracted as a value-unit pair (e.g., $\mu M$, $\text{s}^{-1}$,\%). Importantly, the system includes a fuzzy header matching strategy, where table columns are aligned to target schema fields based on string similarity when exact matches are absent. This enables robust extraction from heterogeneous table formats with noisy or domain-specific headers.

An overview of the entire extraction workflow—including PDF processing, OCR, entity recognition, relation alignment, and structured output formatting—is illustrated in Figure~\ref{fig_tableGet}. In addition to the conventional deep learning-based components, our pipeline is designed in a modular fashion that allows seamless integration of LLMs for document understanding and semantic parsing\cite{huang2024critical_chatgpt}. This design enables flexible substitution or augmentation of key components (e.g., relation extraction or schema alignment) with prompt-driven LLM modules, providing a scalable framework adaptable to heterogeneous document types and future model advances.

To assess the reliability of the automatic extraction, we manually reviewed approximately 20\% of the extracted tables (54 tables in total), which were independently checked by two graduate-level annotators with expertise in enzymology. After cleaning and validation, we retained 175 high-quality tables, resulting in a total of 612 structured reaction entries suitable for downstream modeling and analysis.
\subsection{Enhancing  Fidelity of Data Extraction through Manual Curation}

Through table recognition, molecular reaction extraction, and semantic interpretation powered by LLMs, the system generates preliminary structured data from scientific documents. This initial output is subsequently reviewed and refined through manual curation. The curated data are then fed back into the AI models and algorithms to support continuous training and iterative optimization. As the annotation process advances, the active learning module and AI-assisted labeling system dynamically adapt to new data, progressively streamlining the processing pipeline and improving annotation accuracy\cite{monarch2021human}.

Figure \ref{fig_client} illustrates our literature annotation system. The task management interface provides a clear overview of the real-time progress for each annotation task. Annotators can inspect and refine the data extracted by our models, with convenient options to modify, delete, or reprocess individual entries. Both pre- and post-annotation data are systematically retained to support model improvement and performance evaluation. These records enable continuous refinement of model accuracy and ensure that the deployed system consistently uses the most semantically accurate and structurally reliable models.

\subsection{Model reconstruction based on a manually annotated dataset}

\begin{figure*}[htbp]
    \centering
    \includegraphics[width=\linewidth]{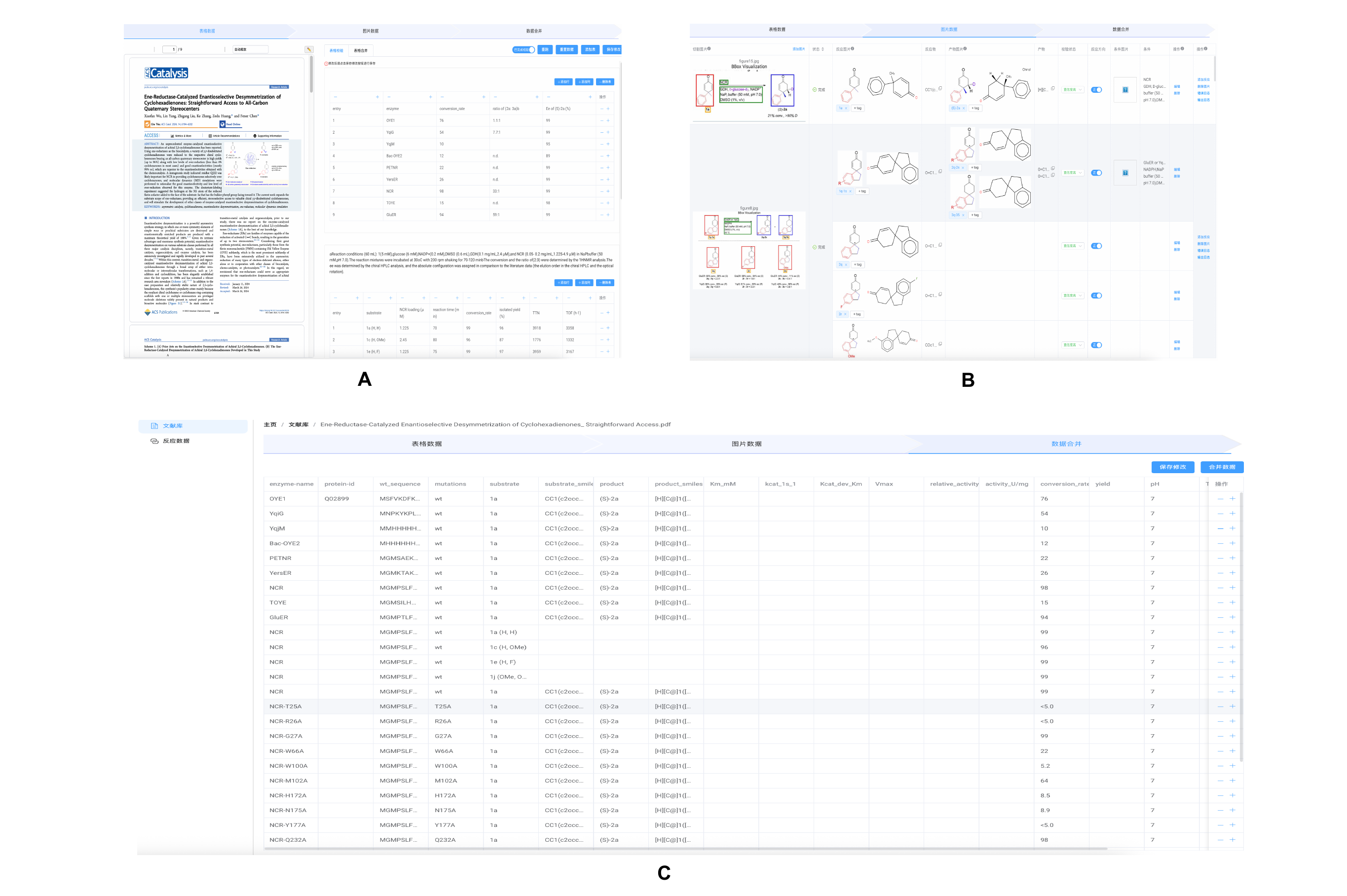}
    \caption{{Interactive platform for multimodal extraction and standardization of enzyme-catalyzed reaction data.}(A) PDF document view with automatic table extraction and region detection. Table content is parsed and aligned with the visual layout.
    (B) Molecular entity and relationship extraction interface. Reactants, products, and intermediate molecules are parsed and linked across chemical reaction records.
    (C) Structured data integration view. Tabular results from enzymatic parameters and molecular relations are merged into a unified schema with fields such as enzyme name, substrate SMILES, mutations, $K_m$, $k_\text{cat}$, and relative activity.
    }
    \label{fig_client}
\end{figure*}

To support large-scale annotation, inspection, and schema-constrained integration of enzymatic reaction data, we developed a modular annotation and extraction system. Figure~\ref{fig_client} illustrates the interface of this system, which includes three coordinated modules:
(A) a document viewing and table annotation panel, allowing users to inspect and align PDF tables in context;
(B) a chemical structure and reaction relationship interface, supporting molecule recognition and cross-reaction linkage;
and (C) a unified structured output module, which aggregates biological fields (e.g., enzyme name, SMILES, $k_\text{cat}$) into a consistent format for model training.
This system was instrumental in building the manually annotated dataset used for model refinement and supports future human-in-the-loop corrections during model deployment.

To improve the adaptability and robustness of the extraction pipeline, we implemented a feedback-driven model reconstruction mechanism based on a manually curated dataset. Although the initial pipeline achieved promising results, it struggled with highly heterogeneous table formats and ambiguous biological field expressions commonly found in enzymology literature. To address this limitation, we conducted large-scale manual annotation on the full set of 270 curated PDFs, resulting in over 1,000 fully labeled tables and approximately 5,000 domain-specific fields, such as $k_\text{cat}$, $K_m$, substrate names, and mutation records. The annotation was carried out by a team of ten researchers with biological backgrounds, following a unified schema-aligned labeling guideline.

The annotated dataset was used to fine-tune both components of the extraction system. Specifically, we retrained the SLANet model to improve structural and semantic field alignment across noisy or nested table layouts. Additionally, a LLM was fine-tuned to better handle reaction-level inference and ambiguous entity resolution. 
This human-in-the-loop strategy enables our system to evolve over time by continually integrating corrections from expert supervision, forming a modular and self-improving framework.

\section{Results}
\subsection{Data preparation}

\subsubsection{Data Preparation for Molecular Graph Recognition}
The "image-to-SMILES" model used in this platform was trained on publicly available datasets, consistent with methodologies established in previous studies\cite{staker2019molecular,yoo2022image,qian2023molscribe,rajan2023decimer}.

To train our molecular reaction recognition models, we prepared a diverse dataset comprising both publicly available sources and our manually annotated images. Specifically, our dataset construction involves two parts:

\noindent\textbf{Publicly Available Datasets:}
Following established benchmark protocols used by other state-of-the-art models, we collected two publicly available datasets: (a) Synthetic Data: approximately 1 million molecular images automatically rendered using theIndigo toolkit\cite{pavlov2011indigo} from molecules randomly sampled from PubChem database\cite{kim2016pubchem}; and (b) Patent Data: around 680,000 molecular images extracted from patents published by the United States Patent and Trademark Office (USPTO)
, accompanied by molecular structure labels.

\noindent\textbf{Manually Annotated Dataset:}
To complement publicly available datasets and enhance model performance on realistic and complex scenarios, we created an in-house annotated dataset. This dataset includes 1,355 manually labeled molecular reaction images, derived directly from scientific literature PDFs in enzymology and chemistry. Each image in this dataset was carefully labeled with the corresponding SMILES representation by experts proficient in chemical structure identification.

\subsubsection{Tabular Data Extraction and Annotation}

For the development and evaluation of our table extraction module, we curated a comprehensive dataset from 270 public-accessible scientific articles published between 2000 and 2024. These articles were filtered using targeted keywords relevant to enzyme kinetics, such as ``P450'' and “enzyme kinetic parameter”. After rigorous screening, we manually annotated over 800 tables, focusing on those containing enzyme-catalyzed reaction data, kinetic parameters and relative enzyme activities.
Each table was reviewed through a dual-phase process, including both initial annotation and subsequent cross-checking by independent annotators. During the review phase, we also evaluated the confidence level for each annotation, and ambiguous cases were flagged for further inspection (as illustrated in Figure~\ref{fig_client}).The key fields are listed in Table~\ref{tab_tabForm}.

To train our table extraction model, we adopted SLANet~\cite{li2022pp} as the backbone architecture and performed a two-stage training process. In the first stage, SLANet was pretrained on the PubTabNet dataset \cite{zhong2019publaynet}, which contains 500,777 training images, 9,115 validation images, and 9,138 testing images, generated by aligning the XML and PDF representations of scientific articles. This large-scale dataset provides high-quality supervision for learning general table structures. In the second stage, we fine-tuned the pretrained model using a domain-specific dataset automatically annotated by our system, which includes tables extracted from biochemical reaction documents. This progressive training strategy allows the model to capture both general structural patterns and domain-specific layout characteristics, thereby enhancing its adaptability and performance in specialized scientific contexts.

\begin{table*}[htbp]
  \caption{Schema for Structured Extraction of Enzyme-Catalyzed Reaction Data}
  \label{tab_tabForm}
  \setstretch{1.0}    
  \renewcommand{\arraystretch}{1.3} 
  \setlength{\tabcolsep}{0pt}       
  \small
  {
  \begin{tabular*}{\linewidth}{
      @{\extracolsep{\fill}}
      >{\raggedright\arraybackslash}p{3.2cm}@{\hspace{1pt}}
      >{\raggedright\arraybackslash}p{6.5cm}@{\hspace{1pt}}
      >{\raggedright\arraybackslash}p{3cm}@{\hspace{1pt}}
      @{}
  }
      \toprule
      \textbf{ Field Name} & \textbf{Description} & \textbf{Data Type}\\
\midrule

      Enzyme name & Name of the enzyme(s) & string \\

Substrate & Chemical substrate used in the reaction & string \\

Enzyme ID & UniProt, GenBank, or PDB ID & string \\

Product & Main product or reaction outcome & string\\

pH & pH value of the reaction & float \\

$K_\text{m}$ & Michaelis constant (value and unit) & object \\

$k_\text{cat}$ & Catalytic constant (value and unit) & object \\

Mutations & Mutation information (list or "None") & list of string / string\\ 


Source Organism & Species of enzyme origin & string \\

Relative Activity & Value and unit; includes yield/conversion \% & object\\ 

      \bottomrule
  \end{tabular*}
  }

\end{table*}

\begin{table*}[htbp]
    \captionsetup{justification=raggedright, singlelinecheck=false}
    \centering
    \begin{threeparttable}
    \caption{Benchmark Comparison of Molecular Image Recognition Models on Synthetic and Realistic Datasets}
    \label{tab_molOcr}
    \renewcommand{\arraystretch}{1.2} 
    \setlength{\tabcolsep}{0pt}       
    \small
    {
    \begin{tabular*}{\linewidth}{
        @{}
        >{\raggedright\arraybackslash}p{1.5cm}@{\hspace{5pt}}
        >{\centering\arraybackslash}p{1.2cm}@{\hspace{5pt}}
        >{\centering\arraybackslash}p{1.5cm}@{\hspace{5pt}}
        >{\centering\arraybackslash}p{1.2cm}@{\hspace{5pt}}
        >{\centering\arraybackslash}p{1.2cm}@{\hspace{5pt}}
        >{\centering\arraybackslash}p{1.2cm}@{\hspace{5pt}}
        >{\centering\arraybackslash}p{1.2cm}@{\hspace{5pt}}
        >{\centering\arraybackslash}p{1.2cm}@{\hspace{5pt}}
        >{\centering\arraybackslash}p{1.2cm}@{\hspace{12pt}}
        @{}
    }
        \toprule
        \multirow{2}{*}{\textbf{Model}} 
     &\multicolumn{2}{c}{\textbf{Synthetic}} & \multicolumn{6}{c}{\textbf{Realistic}} 
    \\
    \cmidrule(lr){2-3}  
    \cmidrule(lr){4-9}  
&Indigo&ChemDraw&CLEF&Staker&ACS&USPTO&JPO&UOB\\
\midrule
MolVec & 95.4&87.9&82.8&0.8&47.4&88.4&67.8&80.6\\
OSRA & 95.0 & 87.3 & 84.6&0.0& 55.3&87.4&55.3&78.5\\
Img2Mol&58.9&46.4&18.3&17.0&23.0&26.3&16.4&68.7\\
DECIMER&69.6&86.1&62.7&40.8&46.5&41.1&55.2&88.2\\
MolScribe&96.0&94.4&89.7&87.8      &69.4&93.6&47.0&96.0\\
\textbf{Ours}&98.9&98.1&95.7&88.7&75.8&94.2&68.8&99.1\\
        
        \bottomrule
    \end{tabular*}
    }
\end{threeparttable}

\end{table*}

\subsection{Table Extraction Accuracy}

\begin{table}[htbp]
    \caption{Performance Comparison on PubTabNet with State-of-the-Art Methods}
    \label{tab_getAcc}
    \setstretch{1.0}    
    \renewcommand{\arraystretch}{1.3} 
    \setlength{\tabcolsep}{0pt}       
    \small
    {
    \begin{tabular*}{\linewidth}{
        @{\extracolsep{\fill}}
        >{\raggedright\arraybackslash}p{2.2cm}@{\hspace{2pt}}
        >{\raggedright\arraybackslash}p{2.3cm}@{\hspace{2pt}}
        >{\raggedright\arraybackslash}p{2cm}@{\hspace{2pt}}
        @{}
    }
        \toprule
        \textbf{Method} & \textbf{Acc(\%)} & \textbf{Gain}   \\
        \midrule
        TableMaster\cite{ye2021pingan} & 77.90\textsuperscript{*}&-  \\
        LGPMA\cite{qiao2021lgpma}& 65.74\textsuperscript{*}&-\\
        SLANet\cite{li2022pp} &86.0 &- \\
        \textbf{Ours}&89.9&\textbf{3.9\%}\\
        \bottomrule
    \end{tabular*}
    }
    \begin{tablenotes}[flushleft]
          \footnotesize
          \item \textsuperscript{*}Source: adapted from Ref. \cite{li2022pp}
      \end{tablenotes}
\end{table}

To evaluate the performance of our table extraction pipeline, we constructed a manually annotated benchmark dataset comprising 270 full-text articles, containing a total of 100 enzyme-related tables. The evaluation metric used was accuracy (Acc), which measures the structural and semantic consistency between the predicted table and the ground truth annotation. As shown in Table~\ref{tab_getAcc}, our approach achieves an accuracy of 89.9\%, outperforming previous state-of-the-art methods such as SLANet (86.0\%), TableMaster (77.90\%), and LGPMA (65.74\%). Compared to the strongest baseline (SLANet), our model achieves a gain of 3.9\%, demonstrating its superior ability to recover biologically relevant tabular structures. The improvement is attributed to the integration of domain-specific schema alignment and post-OCR normalization strategies.

\subsection{ Molecular Graph Recognition and Relation}

\afterpage{\clearpage}
We first evaluated the performance of our molecular graph recognition module, which aims to convert chemical structure images into standardized molecular representations for subsequent reaction analysis. The task was benchmarked against five representative baselines, including MolVec (https://molvec.ncats.io), OSRA\cite{smolov2011imago_OSRA}, Img2Mol\cite{clevert2021img2mol}, DECIMER\cite{rajan2023decimer}, and MolScribe\cite{qian2023molscribe}, on both synthetic and realistic datasets. As shown in Table~\ref{tab_tabForm}, our model consistently outperforms all other methods across all datasets.
On the synthetic datasets generated from Indigo and ChemDraw templates, our method achieves accuracies of 98.9\% and 98.1\%, respectively, surpassing MolScribe, the second-best performer, by 2.9\% and 3.7\%. The performance gap becomes even more significant on realistic datasets (e.g., CLEF, USPTO, and UOB), where model robustness to noise, font variance, and layout distortion is crucial. Our model achieves 95.7\% on CLEF and 99.1\% on UOB, representing absolute improvements of 6.0\% and 3.1\% over the best baseline, respectively. This demonstrates the effectiveness of our model not only in clean settings but also under real-world conditions.

In addition to chemical image-to-SMILES recognition, we also evaluated the relation extraction component, which associates recognized molecular graphs with their roles in enzyme-catalyzed reactions (e.g., substrate, product, cofactor). The module benefits from visual-semantic alignment and rule-guided decoding, achieving a role assignment accuracy of 94.2\% on the manually annotated reaction benchmark\cite{fan2024openchemie}.

These results collectively validate the capability of our system in accurate molecular graph recognition and relation mapping, laying a robust foundation for downstream structured extraction of enzymatic reactions.

\subsection{Model Refinement through Human-Annotated Data}

To evaluate the potential for self-improvement in our extraction framework, we conducted a round of model refinement using a manually annotated dataset. The dataset was constructed from 270 enzymology-related PDFs, yielding over 1,000 labeled tables and 5,000 annotated biological fields including $k_\text{cat}$, $K_m$, relative activities, enzyme names, sequences, substrates, products, mutations as well as other conditions. Annotation was performed by a team of ten domain experts following a standardized schema-aligned guideline.
We used this dataset to fine-tune two critical components of our system. First, we retrained the table structure recognition module based on SLANet\cite{li2022pp}, improving its ability to handle irregular layouts, nested headers, and domain-specific column variations. Second, we fine-tuned the LLM\cite{huang2024critical_chatgpt} responsible for reaction-level parsing, focusing on role assignment (e.g., substrate/product) and molecular relationship inference.
Post-training evaluation showed measurable improvements. The table field extraction accuracy improved by 3.9\%, while molecular relation recognition improved by 3.0\%, as shown in Table~\ref{tab_molOcr}. These results demonstrate the effectiveness of human-in-the-loop feedback for enhancing domain-specific extraction accuracy and confirm the self-evolving nature of our framework\cite{monarch2021human}.
Prior work has demonstrated that thoughtful dataset curation can significantly enhance downstream prediction performance~\cite{meng2022boosting}, reinforcing the importance of our expert-annotated dataset in improving extraction accuracy.

\section{Conclusion}
zERExtractor represents a significant advancement in automating the extraction of enzyme-catalyzed reaction data from scientific literature. By integrating domain-adapted deep learning with LLMs in a modular pipeline, it achieves superior performance in table recognition (89.9\% accuracy), molecular image interpretation (up to 99.1\%), and enzymetic reaction extraction. The system's adaptive learning mechanism, facilitated by expert-guided validation and active learning, enables continuous improvement across diverse data sources. Additionally, the release of a large-scale, expert-annotated P450 enzymes benchmark dataset provides a valuable resource for the enzyme informatics community. Collectively, these innovations bridge the data gap in enzyme kinetics and lay a robust foundation for future deep learning-powered enzyme modeling and biochemical knowledge discovery.

\section{Acknowledgment}\label{acknowledgment}
This work was partly supported by the National Key Research and Development Program of China under Grant No. 2023YFA0915500 (S.W., G.Z. and L.Z.) and 2024YFA0919702 (Y.W.), National Science Foundation of China under grant no. 62272449 (Y.W.) and 12426303 (Y.W.), the Shenzhen Basic Research Fund under grant no KQTD20200820113106007, ZDSYS20220422103800001. We would also like to thank the funding support by the Key Laboratory of Quantitative Synthetic Biology, Chinese Academy of Sciences under grant no. CKL075. This research is also supported by the Science and Technology Development Fund of Macau (0004/2025/RIA1 to J.G.).

\FloatBarrier 



\bibliography{refs}
\end{document}